# Precision psychiatry: predicting predictability

Edwin van Dellen[1,2]

[1] *Department of Psychiatry and University Medical Center Utrecht Brain Center, Utrecht University, Utrecht, the Netherlands.*
[2] *Department of Neurology, UZ Brussel and Vrije Universiteit Brussel, Brussels, Belgium*

**Correspondence**
Edwin van Dellen, MD, PhD
Department of Psychiatry
University Medical Center Utrecht, Brain Center
Heidelberglaan 100, 3584 CX Utrecht, The Netherlands
Email: e.vandellen@umcutrecht.nl

## Abstract

Precision psychiatry is an ermerging field that aims to provide individualized approaches to mental health care. Multivariate analysis and machine learning are used to create outcome prediction models based on clinical data such as demographics, symptom assessments, genetic information, and brain imaging. While much emphasis has been placed on technical innovation, the complex and varied nature of mental health presents significant challenges to the successful implementation of these models. From this perspective, I review ten challenges in the field of precision psychiatry, including the need for studies on real-world populations and realistic clinical outcome definitions, consideration of treatment-related factors such as placebo effects and non-adherence to prescriptions. Fairness, prospective validation in comparison to current practice and implementation studies of prediction models are other key issues that are currently understudied. A shift is proposed from retrospective studies based on linear and static concepts of disease towards prospective research that considers the importance of contextual factors and the dynamic and complex nature of mental health.

## Introduction

Predicting treatment outcomes and prognosis for psychiatric patients remains a daunting task. Precision psychiatry is a branch of research focused on this problem(Vieta, 2015; Fernandes *et al.*, 2017): it aims to tailor treatments by increasing the predictability of outcomes for individual patients(Bzdok *et al.*, 2021). The approach is inspired by precision medicine research based on data-driven analyses; machine learning algorithms are trained on multiple variables to make diagnostic classifications or predictions. The question arises when we can reap the benefits from prediction algorithms in clinical practice(Chekroud *et al.*, n.d.; Stein *et al.*, 2022)?

Methodologically, the precision approach builds on the foundation of statistical prediction models. In the 1950s, Paul Meehl questioned clinicians' ability to make predictions based on their clinical assessments. He posed that statistical predictions outperform clinical judgements when it comes to diagnosis and treatment indication(Meehl, 1956). However, the integration of clinical assessments of an individual with group-level statistical information remained an unsolved problem. The first attempts to solve this issue with artificial intelligence date back to the 1970s, when so-called expert systems were introduced. Expert systems were computer programs that were assigned the task to mimic human decision-making, including clinical decisions(Kassirer and Gorry, 1978). Although promising at the time, this work failed to transform clinical practice. The interest in biological psychiatry later shifted towards biomarker studies and biological subtyping(Kapur *et al.*, 2012). With advances in machine learning methodology in the last decade, and success of precision medicine approaches in other fields such as oncology, precision psychiatry, gained interest. It has the advantage over the expert systems from the seventies that the technology is more sophisticated, while big datasets containing a range of information sources are now available (e.g., medical records, genetics, neurophysiology and neuroimaging data). Additionally, these data-driven approaches may shed new light on pathophysiological pathways(Bzdok and Meyer-Lindenberg, 2018; Grzenda *et al.*, 2021). Several recent studies report promising results(Chekroud *et al.*, n.d.; Williams, 2016; Fernandes *et al.*, 2017; Dwyer *et al.*, 2018; Meehan *et al.*, 2022), for example by predicting antipsychotic treatment response and side-effects with high accuracy(Koutsouleris *et al.*, 2016; Coutts *et al.*, 2023; Dominicus *et al.*, 2023). Unfortunately, a closer look at the data currently used for such studies from a clinical point of view suggests that the desired clinical breakthrough is far from within reach. In this perspective, ten clinical and statistical issues in the precision psychiatry literature are discussed. Examples used in this paper mainly focus on schizophrenia spectrum disorders because this is the most studied population in the precision psychiatry, but topics discussed here are generalizable to order disorders. Some issues for model development based on retrospective datasets from clinical trials are identified, and promising ways forward are highlighted to really make the translation to clinical implementation.

1. Classification or prediction: precisely what?

Prediction of future outcomes is the most clinically relevant application of the precision approach. Data-driven classification of patients compared to a 'gold standard' such as the Diagnostic and Statistical Manual of Mental Disorders (DSM)(American Psychiatric Association, 2013) is of little value because the specific mix of an individual's symptoms and their evolution over time often poorly fit into one classification(van Os *et al.*, 2000; Plana-Ripoll *et al.*, 2019; Romero *et al.*, 2022). Heterogeneity in symptoms also exists between patients with the same classification, the classification itself is a poor indicator for treatment susceptibility, and while some possible pathophysiological associations have been identified, these do not form the basis of the diagnosis as they invariably have low diagnostic likelihood ratio's (van Os *et al.*, 2019a).

Precision studies therefore focus on data-driven subtyping of patients based on existing datasets, subsequently comparing the prognosis or treatment susceptibility between subtypes. Alternatively, retrospective studies may attempt to predict outcomes using data from completed treatment trials; prediction models based on randomized controlled trials (RCTs) often use data from the active treatment group to identify patient characteristics that may predict response. However, this approach also has several potential pitfalls that limit translation to clinical practice, as will be discussed in the following.

## 2. Patient selection

Patients with psychiatric disorders described in the scientific literature on treatment response and/or prognosis were mostly required to give informed consent for study participation, and for good reasons. However, patients with certain characteristics, for example those who are severely paranoid at time of assessment, are systematically undersampled as a consequence(Taipale *et al.*, 2022). Similarly, patients are often excluded if treated under justiciary coercive measures(Luciano *et al.*, 2014). Studies based on these data will thus consider, on average, moderately ill patients(Taipale *et al.*, 2022). This is a well-known limitation of clinical trials for the generalizability of findings to other populations and settings, such as patients with severe psychosis. When clinical information is used as input variable for a prediction model of e.g., treatment outcome, this selection has additional negative impact: the (distribution of) input information deviates from the data in real-world clinical settings, further reducing the generalizability of findings(Brand *et al.*, 2022). For psychosis treatment, male sex, unmet psychosocial needs and functional deficits are examples of predictors of worse clinical outcome that also increase the likelihood of coercive measures being applied(Koutsouleris *et al.*, 2016). As coercive measures are often an exclusion criterion of clinical trials, this will negatively impact prediction model performance in clinical practice.

Future studies should therefore train models based on real-world data with limited exclusion criteria where possible. Data harmonization initiatives that are currently being developed are crucial to ensure that naturalistic data are of sufficient quality to make generalizable inferences(Research Harmonisation Award | Schizophrenia International Research Society, n.d.).

## 3. Fairness

Diversity and inclusion are essential to consider in precision medicine approaches. This is particularly relevant in the field of psychiatry, as societal exclusion and discrimination are directly linked to the development of psychiatric disorders. In the machine learning field, inclusion is closely related to the concept of fairness, which refers to the idea that machine learning models should not be biased or discriminatory(Mitchell *et al.*, 2021). To address fairness, one approach is to ensure that the algorithms themselves are not biased or based on discriminatory variables (algorithmic fairness). Another approach is to consider the impact of the model on different groups of people (group fairness). For example, non-native speakers may have been excluded from studies because standardized interviews are otherwise not available, or data have been obtained in psychiatric hospitals which are less accessible for specific groups due to insurance discrimination(Mamun *et al.*, 2019). Finally, individual fairness involves treating individual instances of data equally. By ensuring that precision models are fair and unbiased, we can use them ethically and responsibly. There may be unresolved or unidentified issues related to diversity and inclusivity in precision psychiatry research. To address these issues and promote an inclusive approach, it is recommendable to include a diversity and fairness statement in precision psychiatry papers for transparency.

### 4. Treatment dose and duration

A substantial number of medication trials treated patients with a dose or duration that is insufficient for the evaluation of treatment efficacy(Howes *et al.*, 2017). Many clinical trials were designed to demonstrate the efficacy of an agent rather than to determine the optimal dose and duration of treatment. Importantly, the optimal dose and minimal treatment duration to reach an effect may vary across subjects, while the optimal dose for treatment effects may often not be reached due to intolerable side effects(Leucht *et al.*, 2013; Zhu *et al.*, 2017; Kahn *et al.*, 2018a). Treatment tolerability is a very important but different issue than treatment effectiveness. Patients can therefore be labeled non-responders to a treatment that is in fact potentially effective because the minimally effective dose is never reached due to intolerability. Finally, many trials have a relatively short follow-up. This may lead to underestimations of the effectiveness (and overestimations of tolerability) of the treatment because a longer follow-up was needed. It may also lead to overestimation of the effectiveness in others, because the treatment effects were only evaluated under strict conditions (e.g., during hospital admission), that may not represent real-world functioning of the patient (figure 2). Note that while these issues are addressed here for medication trials, similar issues can occur in studies of other interventions such as psychotherapy or brain stimulation. Minimally effective dose and duration should therefore be defined in outcome prediction studies but are currently rarely reported, and personalized estimates of dose and duration appropriateness should be obtained in prospective studies where possible.

### 5. Treatment response

A major limitation of retrospective prediction studies on clinical trial data are the lack of consideration of the placebo effect, Hawthorne effect (the phenomenon where people modify their behavior and may experience symptom reduction due to the fact that they are being observed or studied) and the natural course of the disorder(Howick *et al.*, 2013). Psychotropic medication or psychotherapeutic effects are likely at least partially based on separate (biological) mechanisms(Chopra *et al.*, 2021). In psychiatry, placebo-effects are relatively stronger than active treatment effects(Leucht *et al.*, 2017; van Os *et al.*, 2019a). For precision psychiatry studies aiming to predict treatment response, especially when based on biological data, this becomes a major problem.

A thought experiment of a study with a theoretical 'perfect predictor' shows the implications of placebo-induced bias. A *perfect predictor* will only label responders due to *active treatment* effects with a deviant prediction score, while all other patients will be labeled non-responder. If this predictor is truly specific to active treatment effects, this means that it will categorize 'placebo-responders' as non-responders: in these patients, there is no relationship between active treatment and reduction of symptoms.

According to an American Psychiatric Association (APA) consensus statement for (neuroimaging) markers, a biomarker must be at least 80% sensitive, 80% specific, and 80% accurate in order to be considered reliable(First *et al.*, 2012). For a perfectly reliable predictor to meet these requirements – be it a biomarker or a predictor of any other nature - a treatment would need to be at least four times (80%/20%) more effective than placebo in order account for placebo response in the 'gold standard' data.

This level of effectiveness is far from reality for psychiatric treatments. For example, 51% of patients suffering from psychosis are estimated to show minimal response to antipsychotic treatment, in comparison to 30% to placebo treatment(Leucht *et al.*, 2017). Thus, for every 51 patients classified as a responder, 30 may have recovered due to effects unrelated to the pharmacological antipsychotic treatment response (labeled false negatives). As a result, the sensitivity of our predictor will drop to 41% (21 true positives out of 51 responders) in the trial, and its accuracy will be 70% (21 true positives + 49 true negatives), failing the APA requirements (figure 1).

Setting a more stringent threshold for treatment response (which could be done because this threshold is arbitrary, as will be discussed later) cannot help to overcome this problem. In antipsychotic treatment trials, the response-ratio between active treatment (23%) and placebo treatment (14%) for 50% symptom reduction was similar to that for minimal response (defined as 20% symptom reduction)(Leucht *et al.*, 2017). With this more stringent threshold for response, sensitivity will even drop to 39%.

To summarize: in psychiatric treatment conditions where placebo effects and natural course of the disorder cannot be disentangled at the individual level, any theoretically perfect predictor will fail the reliability test in clinical trials. Studies reporting predictors of treatment response with high performance levels without accounting for these issues should caution readers that the reliability of the model may be overestimated.

## 6. Treatment non-response

It may be argued that the effects of placebo and natural fluctuations in mental health can be circumvented by making non-response instead of response the target of our outcome predictor. However, several factors may cause false negatives (i.e., treatment is labeled ineffective for a person, even though it could have been beneficial) in the group of non-responders. For example, in patients with schizophrenia spectrum disorders, non-adherence to treatment is approximated at 50% (adherence is here defined as medication taken as described at least 75% of the time)(Lacro *et al.*, 2002). In a study of our perfect predictor, these participants may be classified as responders while they are clinically classified as non-responders, and will therefore be considered as 'false positives'. Even when placebo-effects are not considered, the accuracy in such a study would be around 75% (24 true negatives+51 true positives), again failing the APA criteria. The Treatment Response and Resistance in Psychosis (TRRIP) Working Group made recommendations for adherence monitoring, but excluding non-adhering patients from trials will likely induce selection bias, and, in the best-case scenario, will lead to 72% adherence(Howes *et al.*, 2017).

Social circumstances and external factors such as ongoing exposure to cannabis or (traumatic) stressors during treatment may further contribute to treatment ineffectiveness(Patel *et al.*, 2016; Marsman *et al.*, 2020). In clinical trials, these factors may be considered random noise in comparisons between active and placebo interventions, but this assumption is not necessarily helpful for the validation of outcome prediction models.

Possible ways forward are the additional inclusion placebo-treatment data in prediction studies where ethically defendable and feasible, or to perform open label trials with blinded discontinuation. This would make it possible to predict the proportional improvement due to

'true' treatment effects(Hafliðadóttir *et al.*, n.d.). Similar approaches could be used to incorporate estimates of natural course of the disorder or non-adherence, in order to improve the real-world performance of the model. Another promising approach in patients with relatively stable states of disorder and a focus on short-term treatment effects is incorporation of information from multiple N=1 trials, and subsequent meta-analysis thereof, where the impact of treatment is randomized within an individual(Hendrickson *et al.*, 2020).

## 7. Outcome definitions

Psychiatric disorders such as psychosis form a spectrum or continuum, ranging from chronically disabling illness to brief, transient and non-clinical experiences(Guloksuz and van Os, 2018). The spectrum is expressed at multiple levels, including symptom severity, genetic liability, neuroanatomical correlates, and functional outcomes after a psychotic episode(van Os *et al.*, 2009; Ripke *et al.*, 2014; van Dellen *et al.*, 2016; Guloksuz and van Os, 2018). Clinical translation of these insights remains an unsolved problem. Guidelines for clinical decisions in patients with psychosis are still largely based on research that used the categorical concept of schizophrenia(van Os *et al.*, 2019a). The state-of-the-art consensus criteria for remission after treatment in psychosis research are the Andreassen remission criteria, which are based on a subset of Positive and Negative Symptom Scale (PANSS) items(Andreasen *et al.*, 2005). Patients diagnosed with psychosis may, however, already fulfill the remission criteria at baseline(Kahn *et al.*, 2018b). Alternatively, treatment response may be defined as an (arbitrarily defined) cut-off point in reduction of symptom severity (e.g., 20% reduction on the PANSS)(Howes *et al.*, 2017; Leucht *et al.*, 2017). Recent trial data show that this will roughly result in a 'median split' dichotomization of the sample into treatment responders and non-responders(Kahn *et al.*, 2018b). This approach may help to gain statistical power and contrast, but is unlikely to represent a (biologically or epidemiologically) plausible contrast between patients, as symptom reduction distributions follow a gaussian distribution (figure 3)(MacCallum *et al.*, 2002; Fried *et al.*, 2022). Prediction models of treatment response based on this approach are therefore unlikely to lead to meaningful insights that can be directly implemented in clinical practice. Continuous treatment outcome measures are more realistic and estimating change in symptom severity may be a way forward. Furthermore, absolute reductions rather than relative reductions in symptoms may be used as outcome measures, because treatment may be more effective in patients with more severe symptoms(Furukawa *et al.*, 2015). At another level, outcomes are often defined based on symptom severity scores. Other outcomes – such as social and existential outcomes – are more relevant for patients, and therefore should be prioritized when an algorithm is used to indicate if a treatment would be suitable for the individual(Maj *et al.*, 2021). A possible mismatch between modeled and desired outcome measures should therefore be considered.

## 8. Validation and implementation

External validation of prediction models in independent, naturalistic cohorts across multiple settings is required in order to establish the generalizability of findings. In practice, validation studies rarely use the same methods as the original work they aim to replicate (if attempts to do so are made at all). Moreover, prediction algorithms need to be tested prospectively (and in multiple n=1 studies where possible) before they can be clinically implemented. The current literature not only lacks such rigorous testing, but also lacks comparison of their performance to existing standards of care(Salazar De Pablo *et al.*, 2021). The evaluation of these models based on symptom severity questionnaires may show mismatch with patient outcomes of factors such as treatment tolerability are not taken into account(Chen and Asch, 2017). Prospective validation of prediction models across real-life outcomes and settings are thus crucial but rarely performed.

While a lot of research is devoted to the development of new outcome prediction models, few studies address how these models should be implemented into clinical care (Salazar De Pablo *et al.*, 2021). Factors that may hamper implementation include limited access to data from the local setting, and unfamiliarity with prediction models among practitioners and patients. This risk increases when the complexity of models increases and implications and assumptions of the model become less transparent.

Finally, implementation of prediction models may shape clinical practice, for example by causing a shift in the composition of the patient population. This can in turn impact the validity of the model. Certain treatment options can become more attractive when the outcome is more predictable (for example if potential severe side-effects of treatment can be ruled-out in advance). This will change the population treated with this intervention, as this treatment may be considered earlier in the treatment protocol. Adaptive modeling approaches are therefore required, but this introduces new challenges, for example regarding privacy(Garralda *et al.*, 2019). Federated learning – a learning paradigm to collectively train algorithms in local settings without exchanging the data itself – is an attractive approach to solve such issues, and health care information processing systems should be transformed to facilitate such approaches(McMahan *et al.*, 2017; Rieke *et al.*, 2020).

## 9. Contextual behavioral factors

From a contextual behavioral perspective, mental health emerges from the dynamic interaction between the individual and the environment. For studies aiming to predict treatment outcomes, this means that the effectiveness of interventions may vary within an individual depending on the setting and circumstances in which the intervention is provided. For example, the treatment response of medication may be (non-linearly) influenced by the setting: response to treatment could be different in clinical versus outpatient care settings with or without community treatment facilities in place. Other factors include the system of friends and family surrounding the patient, the local mental health care system (e.g., private versus public insurance systems), concomitant treatments (e.g., pharmacological treatment with or without parallel psychotherapy) and judiciary status (e.g., voluntary or coercive treatment)(Glick *et al.*, 2011; Kessing *et al.*, 2013; Koutsouleris *et al.*, 2016; Polese *et al.*, 2019; Taipale *et al.*, 2022). The fact that the impact of interventions is context-dependent is further illustrated by the increase of placebo response over time in psychiatric clinical trial data(Weimer *et al.*, 2015).

Precision psychiatry (implicitly) assumes that treatment response markers are stable over time and context, which may not be the case; all the factors mentioned above may change over time within an individual. Cultural factors, beliefs, expectations and values of the individual that is to be treated within a precision framework may also contribute to the distress caused by mental health symptoms, both in a positive and negative way(de Andino and de Mamani, 2022). Integrating the variability of mental health and behavior in the (often biologically oriented) precision framework is a major challenge(Köhne and Van Os, 2021). A possible solution is the use of an integrative approach during model development, where static and dynamic factors contributing to outcomes are combined. In addition, quantitative and qualitative research, within the same study sample and in cocreation with patients, may strengthen model validity and may lead to additional insights.

10. From linear predictions to complex dynamics

Taken together, linear prediction models of outcomes in precision studies are unlikely to lead to improvements of clinical care. Even with complex machine learning approaches, the underlying assumption remains that a combination of factors at baseline will linearly lead to a predictable outcome(Van Os and Kohne, 2021). It has also been argued that even successful implementation of precision medicine may only have limited impact from a public health perspective (Joyner and Paneth, 2015). So how to move forward?

There is compelling evidence that mental health is better understood as a complex dynamical system(Fried and Robinaugh, 2020; Borsboom *et al.*, 2022). Complexity theory suggests that systems are unique and should be approached individually. There is rich diversity in clinical symptoms of patients and in the contributing factors to their mental health(Fried *et al.*, n.d.; van Os *et al.*, 2019b). These factors include positive contributing factors in addition to psychiatric vulnerabilities(Huber *et al.*, 2016). All these factors are interconnected in systems, and their interactions influence outcomes(Borsboom, 2017). Advances in psychiatric symptom network analysis are therefore promising, and require further integration with biological, psychological, and social factors.

In complex dynamical systems, the history of individual elements is crucial for the probability distribution of future outcomes. This again contrasts with the idea that outcomes of future patients can be made predictable based on retrospective analysis of data from others. It stresses the importance of prevention in mental health care, and fits naturally in descriptive approaches used in clinical practice when we take patients' personal histories(Psaty *et al.*, 2018). Computational psychiatry and implementations of virtual trials based on personal data, as currently under development in neuroscience, may be important steps forward(Huys *et al.*, 2016; de Haan, 2017). Finally, as survival of dynamical complex (eco)systems depends on their adaptivity or resilience(Gao *et al.*, 2016), specific interventions should go hand in hand with an intervention that increase resilience and flexibility(Davydov *et al.*, 2010).

## Conclusion

By leveraging the advances in technology and the availability of large datasets, precision psychiatry approaches may contribute to the predictability of prognosis and response to prevention or treatment. Future research should consider limitations of currently available datasets including selection bias, fairness, the noisy reality of treatment data from clinical trials, and incorporate contextual behavioral factors in a broader framework of mental health as a complex dynamical system. Research on methodological innovations should consider implementation in the real-world settings early on in the process.

## Acknowledgements

I thank Jim van Os and Arjen Slooter for their insightful comments on an earlier version of this manuscript.

## Financial support

This work was supported by The Netherlands Organization for Health Research and Development (ZonMW) GGZ fellowship, Award ID: 60-63600-98-711, and a Rudolf Magnus Fellowship from the UMC Utrecht Brain Center.

**BOX 1 – Nomenclature in precision psychiatry**

- *Precision psychiatry* - *Precision* in the context of precision medicine refers to similar outcomes with repeated measurements(Ashley, 2016). Interventions may be targeted with more precision when they are based on better characterization of similarities with other patients.
- *Personalized psychiatry* - The term precision psychiatry is sometimes used as an interchangeable term for personalized psychiatry, but they have slightly different meanings. Personalized psychiatry aims to tailor interventions to specific individuals. Precision psychiatry may thus be used to develop models that help to inform patients more accurately about expected outcomes of interventions, and this information can aid personalized clinical decisions(Council, 2011).
- *Biomarker* – A biomarker is a measurable indicator of a biological state or condition. In the context of precision psychiatry, a biomarker could be used as an indicator of treatment response or prognosis.(First *et al.*, 2012)
- *Machine learning* – a form of artificial intelligence where data and algorithms are used to imitate human learning, hereby improving task performance
- *Predictor* – independent variable in a statistical model that contains information about the occurrence of an event
- *Accuracy* - Accuracy refers to the extent to which an outcome reflects the truth. An example is the fraction of correctly predicted outcomes of a prediction model. Accuracy may thus be used to evaluate the merit of precision approaches as compared to a randomized, one-size-fits-all approach.

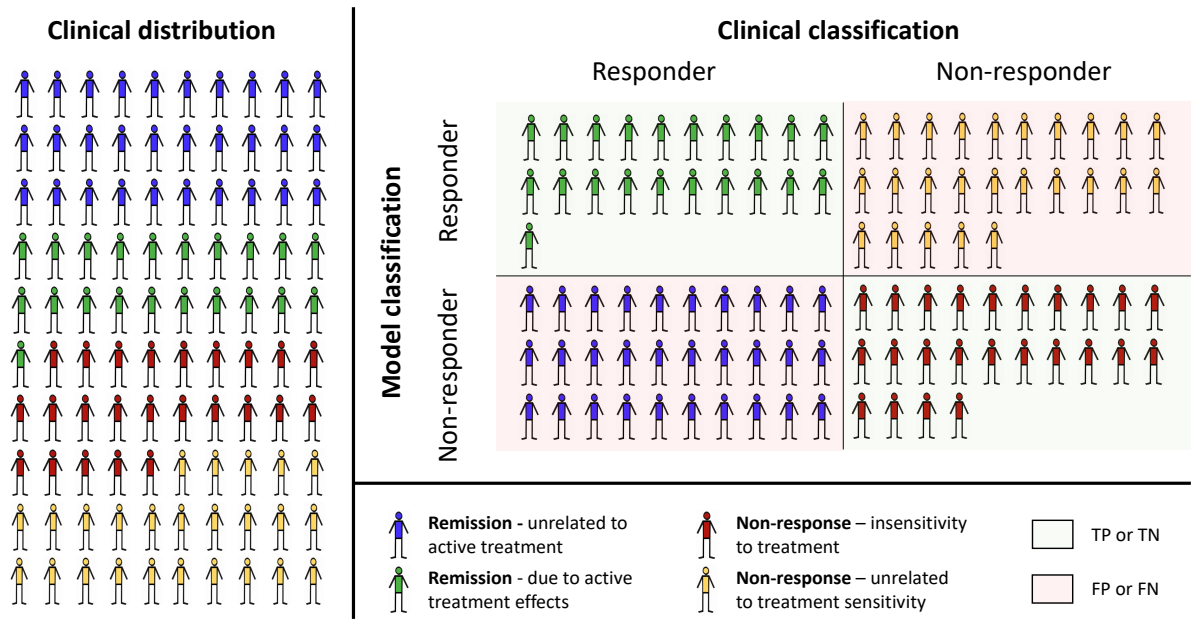

**Figure 1.** Expected model performance for a 'gold standard' model tested on clinical data of patients with schizophrenia spectrum disorders.
*Clinical distribution (left panel) based on* (Lacro *et al.*, 2002; Leucht *et al.*, 2017; Marsman *et al.*, 2020). *In a clinical dataset, for example obtained in a randomized controlled trial of an intervention such as antipsychotic medication, patients are clinically classified as responder or non-responder based on a clinical evaluation at follow-up. Baseline information may be used to predict such outcomes retrospectively, and tested against this clinical classification. This is visualized for a theoretical 'perfect predictor' (right panel), that will have low accuracy in practice. Patients may have received remission due to factors unrelated to the active treatment (e.g., placebo-effects), and meta-analyses suggest this is the case for 30/51 responders. Similarly, non-response may be the result of non-treatment related factors, such as treatment non-adherence or social factors (approx. 25/49 non-responders). As a result, prediction models based on such study designs will have false positive assignments to a response group and false negative assignments to a non-response group. Models based on this approach are therefore unlikely to reach the accuracy needed for implementation in clinical practice. Abbreviations: TP = true positive; TN = true negative; FP = false positive; FN = false negative.*

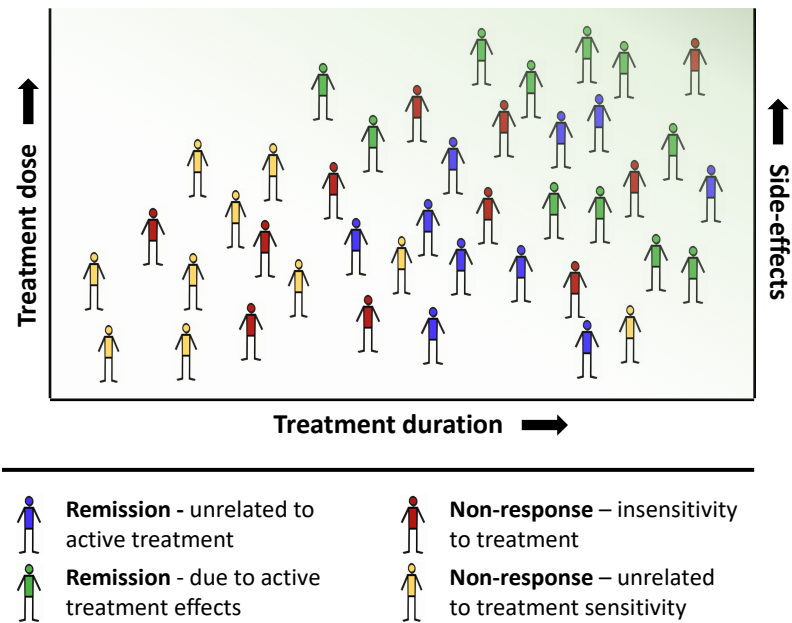

**Figure 2.** Distribution of response and remission classification as a functional of treatment dose and duration.
*Patients treated with medication (or other interventions such as psychotherapy) in treatment response prediction studies are often classified as responder/remitter or non-responder. Treatment dosing and duration however vary in clinical trials, and the chosen regime may lead to inaccurate classifications due to underdosing or too short treatment durations. In addition, patients may withdraw from treatment due to intolerable side effects before reaching an optimal dose for treatment effects. These factors limit the validity of clinical data to be used as 'gold standard' for treatment response prediction.*

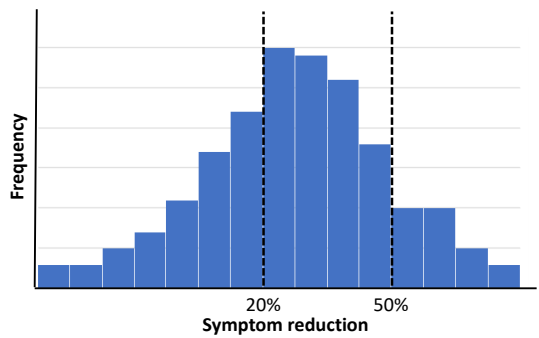

**Figure 3.** Visualization of setting an arbitrary cut-off in symptom reduction on the distribution of responders and non-responders in clinical data.
*Treatment outcome studies often use a relative symptom reduction after treatment with an arbitrary cut-off point (e.g., 20% or 50% reduction compared to the individuals baseline symptom severity score) to define treatment response. The implicit assumption of this approach is that patients can be dichotomized in responders and non-responders. Clinical data from treatment studies however often show a Gaussian distribution in both absolute and relative symptom reduction. As a result, the arbitrary cut-off limits the (pathophysiological) plausibility of such prediction models*(Fried *et al.*, 2022)*. The use of continuous outcomes would therefore be preferable.*